\documentclass[conference]{IEEEtran}
\usepackage[utf8]{inputenc}
\usepackage[T1]{fontenc}
\usepackage{float}
\usepackage{amsmath}
\usepackage{amsfonts}
\usepackage{mathtools}
\usepackage{amsfonts}
\IEEEoverridecommandlockouts
\usepackage{amssymb}
\usepackage{graphicx}
\usepackage{epstopdf}
\usepackage{makeidx}
\usepackage{csquotes}
\usepackage{graphicx}
\usepackage{cite}
\usepackage{xcolor}
\usepackage{subcaption}
\usepackage{graphicx}
\usepackage{balance}

\begin{document}

\title{CR-Enabled NOMA Integrated Non-Terrestrial IoT Networks with Transmissive RIS}
\author{Wali Ullah Khan$^\star$, Zain Ali$^\ddagger$, Asad Mahmood$^\star$, Eva Lagunas$^\star$, Syed Tariq Shah$^\mp$ Symeon Chatzinotas$^\star$ \\$^\star$Interdisciplinary Centre for Security, Reliability and Trust (SnT), University of Luxembourg, Luxembourg\\
$^\ddagger$Electrical and Computer Engineering Department, University
of California, Santa Cruz, USA\\
$^\mp$School of Computer Science and Electronic Engineering, University of Essex, Colchester, UK\\
\{waliullah.khan, asad.mahmood, eva.lagunas, symeon.chatzinotas\}@uni.lu\\
zainalihanan1@gmail.com, syed.shah@essex.ac.uk

\thanks{This work has been supported by the Luxembourg National Research Fund (FNR) under the project MegaLEO (C20/IS/14767486).}
}%
\maketitle

\begin{abstract}
This work proposes a transmissive reconfigurable intelligent surface (T-RIS)-equipped low earth orbit (LEO) satellite communication in cognitive radio-enabled integrated non-terrestrial networks (NTNs). In the proposed system, a geostationary (GEO) satellite operates as a primary network, and a T-RIS-equipped LEO satellite operates as a secondary Internet of Things (IoT) network. The objective is to maximize the sum rate of T-RIS-equipped LEO satellite communication using downlink non-orthogonal multiple access (NOMA) while ensuring the service quality of GEO cellular users. Our framework simultaneously optimizes the total transmit power of LEO, NOMA power allocation for LEO IoT (LIoT) and T-RIS phase shift design subject to the service quality of LIoT and interference temperature to the primary GEO network. To solve the non-convex sum rate maximization problem, we first adopt successive convex approximations to reduce the complexity of the formulated optimization. Then, we divide the problem into two parts, i.e., power allocation of LEO and phase shift design of T-RIS. The power allocation problem is solved using Karush–Kuhn–Tucker conditions, while the phase shift problem is handled by Taylor approximation and semidefinite programming. Numerical results are provided to validate the proposed optimization framework.
\end{abstract}


\IEEEpeerreviewmaketitle

\section{Introduction}
Non-terrestrial networks (NTNs) are gaining attention for their global wireless coverage and diverse service capabilities \cite{9861699}. NTNs consist of satellites in the space layer and unmanned aerial vehicles and high-altitude platforms in the atmospheric layer. We can further classify space into low-earth orbit (LEO), medium-earth orbit (MEO), and geostationary-earth orbit (GEO), where satellites operate at varying heights \cite{10097680}. LEO satellites can orbit 500 km above the earth's surface and offer high data rates, low transmission latency, and ubiquitous connectivity. LEO satellites can connect Internet of Things (IoT) devices across the globe for various applications in smart cities and remote areas. However, one of the main challenges in NTNs will be the efficient integration between different layers and their coexistence with terrestrial networks \cite{9275613}. Another issue can be the efficient reuse of available spectrum and the management of cochannel interference among terrestrial and NTNs \cite{9852737}. Furthermore, advanced satellite systems in NTNs use large antenna arrays with complex RF modules and limited energy resources on board, resulting in high energy consumption \cite{10396846}. Therefore, research efforts are required to explore sustainable and energy-efficient technologies and efficient spectrum allocation methods.

To enhance the channel capacity and improve the spectrum efficiency, reconfigurable intelligent surfaces (RIS) and non-orthogonal multiple access (NOMA) can be promising solutions for NTNs \cite{10508590}. With engineered manipulating capabilities, RIS can modify the phase and amplitude of the wireless signals, which improves the coverage and capacity of NTNs. It can operate in reflective, transmissive, and hybrid modes \cite{10584518}. RIS can also address the issue of noise amplification and the additional noise produced by traditional relay systems \cite{AHMED2024100668}. On the other hand, NOMA empowers the communication system to serve multiple devices over the same spectrum at the same time \cite{9543581}. It can be achieved by superposition coding and successive interference cancellation (SIC) methods. More specifically, the transmitter can allocate different power levels to superimpose the signals of multiple devices over the same frequency, and the receivers can apply SIC to decode their desired signals \cite{9903905}.

Most of the existing literature considers the application of reflective RIS in NTNs. For example, in \cite{9917323}, the authors have maximized the weighted sum rate of integrated terrestrial NTNs assisted by RIS. They adopted an alternating optimization (AO) approach for designing the beamforming vector of BS and the phase response of RIS. Accordingly, Feng {\em et al.} \cite{10494510} have employed the AO method to solve the problem of active beamforming at satellite and phase response at RIS for maximizing the received signal power. Moreover, Ge {\em et al.} \cite{9947334} have proposed a secure cooperative communication in RIS-assisted integrated terrestrial NTNs using the AO technique. In particular, they minimized the transmit power of BS by designing a cooperative beamforming subject to ensure the secrecy of satellite users and the service quality of terrestrial users. Then, Tekbiyik {\em et al.} \cite{9954397} have also utilized RIS to assist terahertz communication among large LEO satellite networks. They investigated the misalignment fading on error performance and signal-to-noise ratio of the system. Further, some works have also studied the energy efficiency optimization in IRS-assisted NTNs \cite{10200793,10365519,9539541,10530613}. More specifically, the authors in \cite{10200793,10365519} have applied the AO approach for NOMA power allocation at the LEO satellite and phase response design at the RIS system. They maximized the energy efficiency of NOMA RIS-assisted LEO communication subject to the service quality of ground users. In \cite{9539541}, authors have investigated the uplink and downlink capacity of the Internet of Things in RIS-assisted satellite communication networks. In \cite{10530613}, Lv {\em et al.} have proposed an OA-based energy efficiency maximization framework by optimizing the transmit power, SIC decoding order, and phase response.

RIS technology has recently been used as a transmitter in terrestrial setups, known as transmissive RIS (T-RIS), without requiring complicated signal processing \cite{9983541}. This distinguishes it from traditional multi-antenna systems that depend on costly RF modules, resulting in expensive hardware expenses. Based on the existing literature on RIS-assisted NTNs, it can be seen that the authors considered reflective RIS to investigate the system performance. To the best of our knowledge, the optimization framework for cognitive radio (CR)-enabled NOMA integrated NTNs with transmissive RIS has not been investigated, and it is an open research topic to investigate. This paper proposes a new optimization framework in CR-enabled NOMA integrated NTNs with T-RIS. In particular, we aim to maximize the sum rate of the secondary T-RIS-equipped LET satellite network while ensuring the service quality of the primary GEO network. The proposed framework simultaneously optimizes the NOMA power allocation at the LEO satellite and phase shift design at T-RIS subject to the interference temperature from the secondary to the primary network and the minimum rate requirement of LEO IoT (LIoT). The problem of sum rate maximization is formulated as non-convex due to rate expressions and coupling optimization variables. Therefore, we first exploit successive convex approximation (SCA) to reduce the overall complexity of the joint problem. Then, we divide the problem into two subproblems, i.e., the power allocation problem and the phase shift design problem and solve it in two steps. In the first step, we calculate the power allocation for the LEO satellite given any random phase shift design at T-RIS using Karush–Kuhn–Tucker (KKT) conditions. Given the optimal power allocation, we compute a phase shift design for T-RIS in the second step. Numerical results are provided to validate the proposed optimization framework.  

\section{System Model}
\begin{figure}[!t]
\centering
\includegraphics [width=0.45\textwidth]{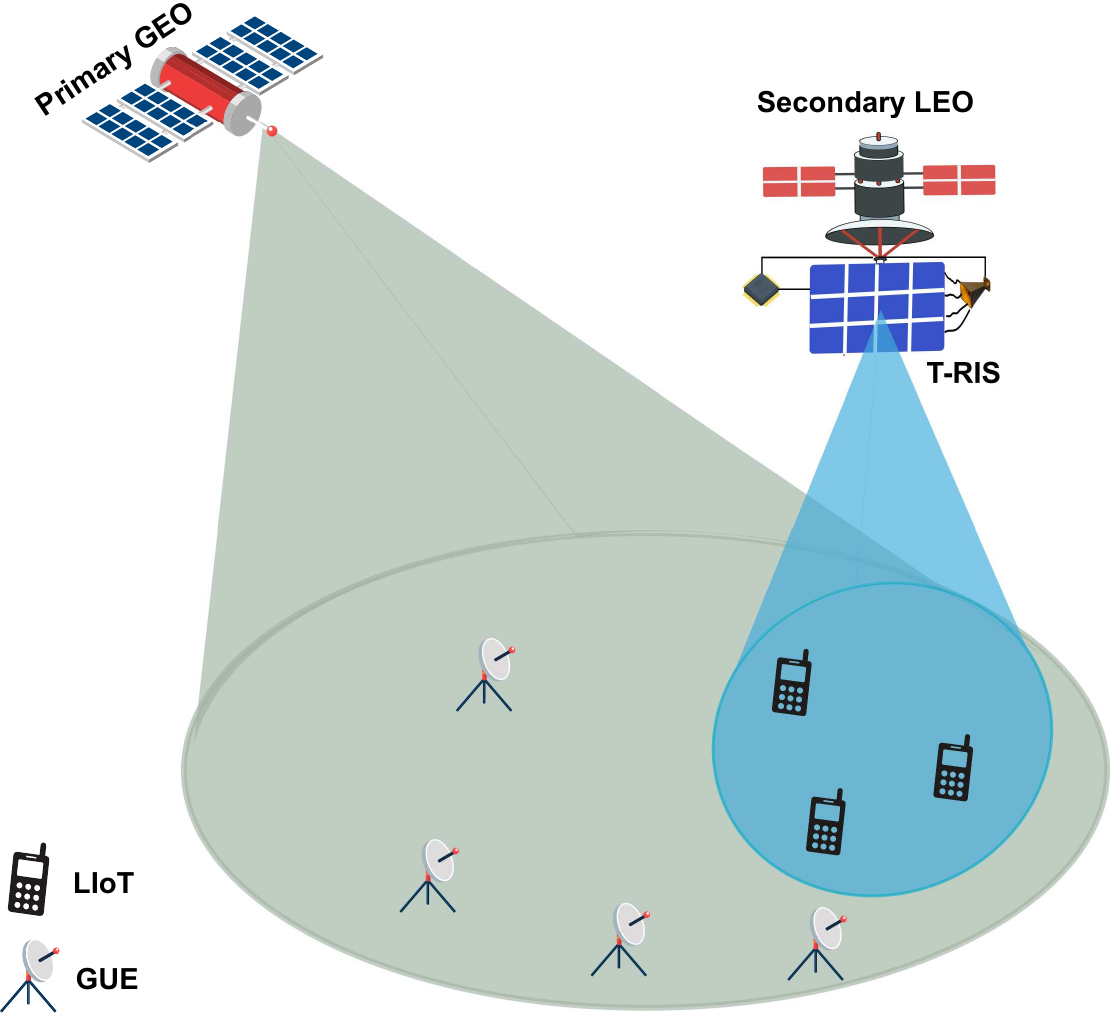}
\caption{System model}
\label{SM}
\end{figure}

This work considers a downlink CR-enabled NOMA integrated NTNs with T-RIS. As illustrated in Fig. \ref{SM}, a GEO satellite operating as a primary network communicates with GEO user equipments (GUEs) using TDMA transmission. At the same time, a T-RIS-equipped LEO satellite, covering the primary network and operating as a secondary IoT network, communicates with LIoT using NOMA transmission. Thus, the primary and secondary networks use the same spectrum and cause co-channel interference. This work maximizes the sum rate of the secondary LEO network while ensuring the quality of services of the primary GEO network. To control the interference from the secondary LEO network to the primary GEO network and ensure the quality of services of GUEs, this work considers a constraint of interference temperature in the proposed optimization framework such that the interference caused by the LEO network should not exceed a predefined threshold. The proposed integrated NTNs assume that the channel state information (CSI) is available in both the primary and the secondary networks. The T-RIS is equipped with LEO through a feed antenna and consists of $M$ elements. Moreover, the $M$ elements constitute a phase shift matrix ${\bf \Phi}_t\in\mathbb C^{M\times M}$, where the beamforming vector can be expressed as ${\bf \Phi}_t=[\eta_1e^{j\phi_1},\eta_2e^{j\phi_2},\dots, \eta_Me^{j\phi_M}]$ with $\eta_m\in\{0,1\}$ is the amplitude and $e^{j\phi_m}$ shows the phase of the transmitted signal. In this work, we consider that T-RIS cannot amplify the signal received from the feed antenna of the LEO satellite and hence $|\eta_m|\leq 1,\ \forall m$. Moreover, the phase constraint of each T-RIS element can be expressed as $\pi_m\in\{0,2\pi\}\ \forall m$.

For efficient SIC decoding at LIoT, we consider that LEO communicates with two LIoT over the same spectrum at a given time. Let us denote $k$ and $j$ are the LIoT, the transmit superimposed signal of LEO for these LIoT can be stated as $x=\sqrt{p_iP_t}x_i+\sqrt{p_jP_t}x_j$, where $p_i$ and $p_j$ are the power allocation coefficients of LIoT $k$ and LIoT $j$ and $P_t$ is the total transmit power of LEO. Furthermore, $x_k$ and $x_j$ are the unit power signals for LIoT $k$ and LIoT $j$. The sum of the power allocation coefficient should be less than/ or equal to one, i.e., $p_k+p_j\leq 1$, according to the downlink NOMA principle. The signal that LIoT $k$ and LIoT $j$ receive from LEO satellite can be written as
\begin{align}
y_k={\bf g}_k\boldsymbol{\Phi}_tx+n_k,\\
y_j={\bf g}_j\boldsymbol{\Phi}_tx+n_j,
\end{align}
where ${\bf g}_k\in\mathcal C^{M\times 1}$ and ${\bf g}_j\in\mathcal C^{M\times 1}$ are the channel vectors between T-RIS-equipped LEO and LIoT $k$ and $j$. Furthermore, $n_k$ and $n_j$ are the additive white Gaussian noise (AWGN) of LIoT $k$ and LIoT $j$. This work adopts a block faded channel model such that ${\bf g}_\kappa=\hat{\bf g}_\kappa e^{\vartheta\pi\psi}$, where $\hat{\bf g}_\kappa$ is the vector of complex-valued channel gains with $\kappa\in\{k,j\}$, $\vartheta$ denotes the imaginary number which is equal to $\sqrt{-1}$ and $\psi$ shows the Doppler shift. Considering the Rician fading, $\hat{\bf g}_\kappa$ can be further expressed as:  
\begin{align}
&\hat{\bf g}_\kappa=[1,e^{-j\rho \sin{\theta}_\kappa\cos{\varphi}_\kappa},\dots, e^{-j\rho \sin{\theta}_\kappa\cos{\varphi}_\kappa(M-1)}]^T
\end{align}
where $\rho=2\pi f_c d_0/c$ such that $c$ is the speed of light, $f_c$ is the carrier frequency, and $d_0$ is the spacing between elements on the T-RIS. Moreover, $\theta$ is the vertical and $\varphi$ is the horizontal angle of departure to $U_\kappa$. For efficient implementation of SIC at the receiver side, we assume that the channel gain of LIoT $k$ is stronger than LIoT $j$. Therefore, LIoT $k$ applies SIC to subtract the signal of LIoT $j$ before decoding its desired signal. However, LIoT $j$ cannot apply SIC and decode the signal by treating the signal of LIoT $k$ as a noise. Based on these observation, the rate of LIoT $k$ and LIoT $j$ can be expressed as $R_k=\log_2(1+\gamma_k)$ and $R_j=\log_2(1+\gamma_j)$. Please note that $\gamma_k$ and $\gamma_j$ are the signal-to-interference plus noise ratios which can be described as:
\begin{align}
\gamma_k=\frac{|{\bf g}_k\boldsymbol{\Phi}_t|^2p_kP_t}{\sigma^2},\label{5}\\
\gamma_{j}=\frac{|{\bf g}_j\boldsymbol{\Phi}_t|^2p_jP_t}{\sigma^2+|{\bf g}_j\boldsymbol{\Phi}_t|^2p_kP_t},\label{6}
\end{align}
where $\sigma^2$ in (\ref{5}) and (\ref{6}) is the variance of AWGN while the second term in the denominator of (\ref{6}) is the NOMA interference from the signal of LIoT $k$. 

During the NOMA transmission, the secondary LEO satellite also causes interference to the GUEs of the primary GEO satellite. To control interference from the secondary LEO network to the primary GEO network and ensure the communication services of GUEs, we invoke an interference temperature constraint in the proposed optimization framework. According to this constraint, the interference caused by secondary LEO satellite to GUEs should be less than or equal to a predefined threshold such as
\begin{align}
{\bf h}_lP_t(p_k+p_j)\leq I_{th}
\end{align}
where $I_{th}$ is the maximum interference threshold and ${\bf h}_l\in\mathcal C^{M\times 1}$ is the channel vector between secondary T-RIS-equipped LEO satellite and GUE $l$ of primary GEO network.
\section{Problem Formulation and Proposed Solution}
This section provides the mathematical problem formulation and solution of the proposed system model in the previous section.
\subsection{Problem Formulation}
This work aims to enhance the performance gain of CR-enabled NOMA integrated NTNs, which can be evaluated in terms of system spectral efficiency. In particular, the proposed framework maximizes the achievable spectral efficiency of secondary NOMA T-RIS-equipped LEO communication by simultaneously optimizing the power allocation of LEO according to the downlink NOMA protocol and phase shift design of T-RIS while ensuring the minimum rate of LIoT. Furthermore, our optimization framework also controls the interference temperature from the secondary network to the primary network in order to ensure the service quality for GUEs. The joint problem of transmitting power and phase shift to maximize the achievable spectral efficiency of the system can be formulated as
\begin{alignat}{2}
\mathcal P_0:\begin{cases}  \underset{{\bf p},\boldsymbol{\Phi}_t,P_t}{\text{max}}\ (R_i+R_j)\label{7}\\
C_1:\ R_{\eta}\geq R_{min},\ \eta\in i,j,\\
C_2:\  h_lP_t\leq I_{th},\\
C_3:\ \phi_k\in\{0,2\pi\},\ k\in K,\\
C_4:\ p_i+p_j= 1,\\
C_5:\ P_t\leq P_{max},
\end{cases}
\end{alignat}
where ${\bf p}\in p_i,p_j$ and $h_l=|{\bf h}_l\boldsymbol{\Phi}_t|^2$ where ${\bf h}_l$ is the channel gain from the T-RIS to the primary receiver. Constraint $C_1$ ensures the minimum rate requirements of LIoT $k$ and LIoT $J$, where $R_{min}$ is used as the minimum threshold; constraint $C_2$ invokes the interference temperature constraint to guarantee the service quality of GUEs in primary network, constraint $C_3$ design phase shift for T-RIS; and constraints $C_4$ and $C_5$ control the LEO transmit power according to the NOMA principle. 
 
The optimization problem $\mathcal P_0$ is non-convex due to the rate expressions in the objective function and constraint $C_1$. Moreover, the problem is also coupled over two variables, i.e., power allocation and phase shift design. To reduce complexity and make the optimization more tractable, we obtain an efficient solution in three steps. In the first step, we adopt SCA which reduces the complexity of the objective function and constraint $C_1$. In the second step, we decouple the problem into separate problems for the power allocation of the LEO satellite and the phase shift of T-RIS. In the third step, for a given phase shift design of T-RIS, we first calculate the closed-form solution of LEO NOMA power allocation. Then, given the optimal power allocation, we design a phase shift of T-RIS. By applying SCA onto $P_0$, the rate expression of LIoT $\eta$ in the objective function and $C_1$ can be re-expressed as

\begin{align}
\Bar{R}_\eta= \alpha_\eta\log_2(\gamma_\eta)+\beta_\eta,
\end{align}

where $\alpha_\eta=\hat{\gamma_\eta}/(1+\hat{\gamma_\eta})$ and $\beta_\eta=\log_2(1+\hat{\gamma_\eta})-\dfrac{\hat{\gamma_\eta}}{(1+\hat{\gamma_\eta})}\log_2(\hat{\gamma_\eta})$, where $\hat{\gamma_\eta}$ denotes the value of $\gamma_\eta$ from the previous iteration. Next, we compute NOMA power allocation at the LEO satellite, given the fixed phase shift design.

\subsection{NOMA Power Allocation}
For any given phase shift design at T-RIS, the problem $\mathcal P_0$ can be simplified into a NOMA power allocation problem at LEO such as
\begin{alignat}{2}
\mathcal P_1:\begin{cases}  \underset{{p_k,p_j,P_t}}{\text{max}}\ (\Bar{R}_k+\Bar{R}_j)\label{7}\\
C_{1.1}:\ \Bar{R}_{\eta}\geq R_{min},\ \eta\in k,j,\\
C_{1.2}:\  h_lP_t\leq I_{th},\\
C_{1.3}:\ p_k+p_j= 1,\\
C_{1.4}:\ P_t\leq P_{max},
\end{cases}
\end{alignat}
The optimization problem $\mathcal P_1$ is the power allocation at the LEO satellite and can be efficiently solved by KKT conditions. First, to tackle $C_{1.3}$ we substitute $p_j=1-p_k$, then, the Lagrangian function is given as:
\begin{align}
&L(p_k,p_j,P_t)=(\Bar{R}_k+\Bar{R}_j)+\lambda_\eta(\Bar{R}_{\eta}-R_{min})\nonumber\\&+\mu_1(I_{th}-{\bf h}_lP_t)+\mu_2(P_{max}-P_t)
\end{align}
where $\lambda_\eta\in\{k,j\},\mu_1,\mu_2,\mu_2$ denote the Lagrangian multipliers associated with the Lagrangian function. Next, we exploit KKT conditions to compute the transmit power of LIoT $k$ first by calculating the partial derivative w.r.t $p_k$, which can be stated as:
\begin{align}
\frac{\partial L(\lambda_\eta,\mu_1,\mu_2)}{\partial p_k}|_{p_k=p_k^*}=0,\label{12}
\end{align}


After calculating the partial derivation, the power allocation coefficient of LIoT $k$ $p^*_k$ can be found as:
\begin{align}
    &p_k^*=\\&\dfrac{(|{\bf g}_j\boldsymbol{\Phi}_t|^2 |{\bf g}_k\boldsymbol{\Phi}_t|^2-|{\bf g}_k\boldsymbol{\Phi}_t|^2 \lambda_k+|{\bf g}_j\boldsymbol{\Phi}_t|^2\lambda_j)\sigma^2}{|{\bf g}_j\boldsymbol{\Phi}_t|^2 |{\bf g}_k\boldsymbol{\Phi}_t|^2 (\lambda_k-\lambda_j) P_t}, \label{pksol}
\end{align}
After getting $p^*_k$ for LIoT $k$, we can efficiently calculate the power of LIoT $j$ as:
\begin{align}
p^*_j=1-p^*_k.
\end{align}
Now it remains to solve the problem for $P_t^*$. As we know the sum rate of the system is a monotonically increasing function of $P_t^*$ it means that the system will allocate as much power for transmission as possible without violating any constraint. Using this fact, the optimal value of $P_t$ is computed by using $C_{1.2}$ and $C_{1.4}$ as upper bounds:
\begin{align}
    P_t^*=\min\left\{\dfrac{I_{th}}{h_l},P_{max}\right\}.
\end{align}
For computation of (\ref{pksol}) we need the values of $\lambda_k$ and $\lambda_j$, in each iteration the values of $\lambda_k$ and $\lambda_j$ are optimized using the sub-gradient method as:
\begin{align}
    &\lambda_\eta= \lambda_\eta+\delta(\Bar{R}_{\eta}\geq R_{min}), \forall \eta=k,j
\end{align}
where $\delta$ is is the step size of the sub-gradient method.
\subsection{T-RIS Phase Shift Design}
Next, given the optimal values of $p^*_k,p^*_j$ at LEO, the problem $\mathcal P_0$ can be efficiently transformed into a phase shift design of T-RIS, which can be written as follows:
\begin{alignat}{2}
\mathcal P_2:\begin{cases}  \underset{\boldsymbol{\Phi}}{\text{max}}\ (\Bar{R}_i+\Bar{R}_j)\label{7}\\
C_{2.1}:\ \Bar{R}_{\eta}\geq R_{min},\ \eta\in i,j,\\
C_{2.2}:\ |{\bf h}_l\boldsymbol{\Phi}_t|^2 P_t \leq I_{th},\\
C_{2.3}:\ \phi_k\in\{0,2\pi\},\ k\in K,
\end{cases}
\end{alignat}
To make the problem tractable we introduce variables $G_k={\bold g}_k {\bold g}_k^\dagger$, $G_j={\bold g}_j {\bold g}_j^\dagger$, $\Phi=\boldsymbol{\Phi}_t \boldsymbol{\Phi}_t^\dagger$ and $H_l={\bold h}_l {\bold h}_l^\dagger$. With these substitutions we get:

\begin{align}
\gamma_k=\frac{\text{tr}(G_k \Phi) p_kP_t}{\sigma^2},\\
\gamma_{j}=\frac{\text{tr}(G_j \Phi) p_jP_t}{\sigma^2+Tr(G_j \Phi)  p_k P_t},
\end{align}
where $\text{tr}$ denotes the trace operator. Then, $\mathcal P_2$ is rewritten as:

\begin{alignat}{2}
\mathcal P_3:\begin{cases}  \underset{\boldsymbol{\Phi}}{\text{max}}\ (\Bar{R}_i+\Bar{R}_j)\label{7}\\
C_{3.1}:\ \Bar{R}_{\eta}\geq R_{min},\ \eta\in i,j,\\
C_{3.2}:\ \text{tr}(H_l \Phi) P_t\leq I_{th},\\
C_{3.3}:\ \Phi \geq 0,\\
C_{3.4}:\ \text{diag}(\Phi) \leq \text{I},\\
C_{3.5}:\ \text{rank}(\Phi) = 1,
\end{cases}
\end{alignat}
where $\text{I}$ denotes the identity matrix, $C_{3.3}$ ensures that $\Phi$ is positive semi-deifinite, $C_{3.4}$ is the amplitude constraint of the RIS elements and $C_{3.5}$ fulfills the requirement that $\Phi$ must have a rank of 1. Further simplification of $\mathcal P_3$ gives us:

\begin{alignat}{2}
\mathcal P_4:\begin{cases}  \underset{\boldsymbol{\Phi}}{\text{max}}\ \log_2( \text{tr}(G_k \Phi) p_k P_t+\sigma^2)-\log_2(\sigma^2) \\ +\log_2(\text{tr}(G_j \Phi) p_j P_t\!\!+\!\!\sigma^2\!\!+\!\!\text{tr}(G_j \Phi) p_k P_t)\\-\log_2(\sigma^2+\text{tr}(G_j \Phi)p_k P_t)\\
C_{4.1}:\ \text{tr}(G_k \Phi) p_k P_t \geq (2^R_{min}-1) \sigma^2,\\
C_{4.2}:\ \text{tr}(G_j \Phi) p_j P_t \geq (2^R_{min}\!\!-\!1) \hat{Q},\\
C_{4.3}:\ \text{tr}(H_l \Phi) P_t\leq I_{th},\\
C_{4.4}:\ \Phi \geq 0,\\
C_{4.5}:\ \text{diag}(\Phi) \leq \text{I},\\
C_{4.6}:\ \text{rank}(\Phi) = 1,
\end{cases}\label{p4}
\end{alignat}
where $\hat{Q}=(\sigma^2\!\!+\!\text{tr}(G_j \Phi) p_k P_t)$. The problem (\ref{p4}) is a non-convex problem due to the $-\log_2(\sigma^2+\text{tr}(G_j \Phi) p_k P_t)$ term in the objective function and the rank-1 constraint. To solve the problem we relax the rank-1 constraint in $c_{4.6}$. Then if the solution violates the rank-1 condition, we employ the Gaussian randomization process to ensure compliance. To tackle the $-\log_2$ term in the objective function, first upper bound the function as:
\begin{align}
    \sigma^2+\text{tr}(G_j \Phi)p_k P_t\leq \Lambda.
\end{align}
This bound transforms the function in the objective as $-\log_2(\Lambda)$. then, we employ Taylor approximation which transforms $\log_2(\Lambda)$ into:
\begin{align*}
    \log_2(\overline{\Lambda})+\dfrac{\Lambda-\overline{\Lambda}}{\overline{\Lambda}}
\end{align*}
where $\overline{\Lambda}$ denotes the value of $\Lambda$ in previous iteration. then the transformed problem becomes:
\begin{alignat}{2}
\mathcal P_5:\begin{cases}  \underset{\boldsymbol{\Phi},\Lambda}{\text{max}}\ \log_2( \text{tr}(G_k \Phi) p_k P_t+\sigma^2)-\log_2(\sigma^2)+\log_2(\text{tr}\nonumber\\ (G_j \Phi) p_j P_t\!+\!\sigma^2\!+\!\text{tr}(G_j \Phi) p_k P_t)\!-\!\log_2(\overline{\Lambda})\!-\!\dfrac{\Lambda\!-\overline{\Lambda}}{\overline{\Lambda}}\\
C_{5.1}:\sigma^2+\text{tr}(G_j \Phi)p_k P_t\leq \Lambda,\\
C_{4.1},C_{4.2},C_{4.3},C_{4.4},C_{4.5},\nonumber
\end{cases}
\end{alignat}
which is a semidefinite programming problem and can be solved using Mosek solver with CVX.
\section{Numerical Results}
In this section, we present the simulation results to evaluate the performance of the proposed framework. For the simulations, the values of the system parameters were taken as $I_{th}$= 2 W, $R_{min}$= 0.1 b/s/Hz, $M$=10, $\sigma^2$=$10^{-7}$.

\begin{figure}[!t]
\centering
\includegraphics [width=0.48\textwidth]{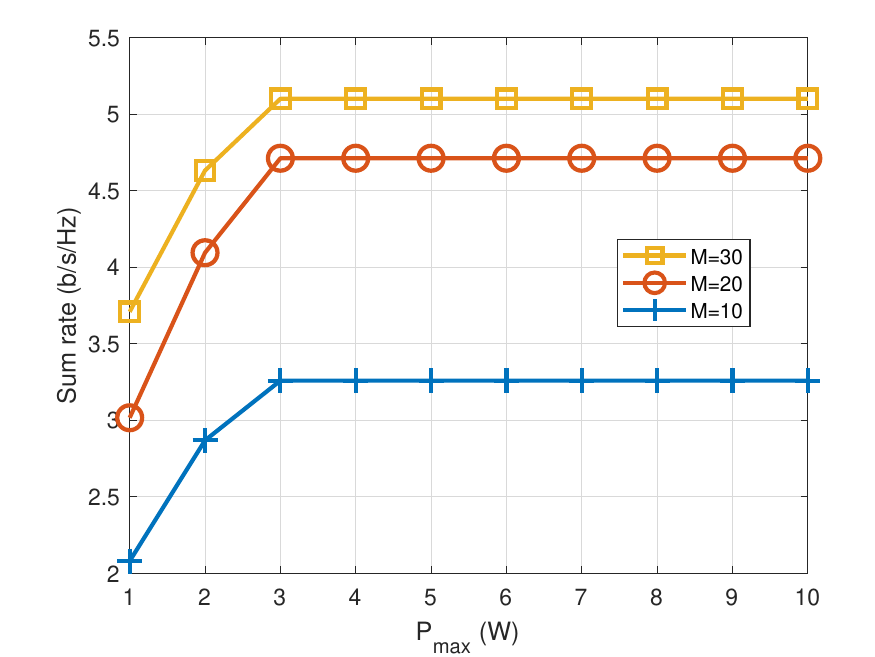}
\caption{Impact of different $P_{max}$ and $M$ on the sum rate of the system.}
\label{fig1}
\end{figure}

The available transmission power plays a vital role in the total achievable rate of a communication network, as more transmission power results in higher channel capacities. Fig. \ref{fig1} shows the impact of increasing the value of $P_{max}$ and $M$ on the sum rate offered by the secondary network. It can be seen that for any value of $M$, increasing $P_{max}$ generally results in an increased sum rate of the system because when $P_{max}$ increases, the LEO can allocate more power for transmission, thereby increasing the sum rate. However, as shown in Fig. \ref{fig1}, after a certain point, further increasing $P_{max}$ has no impact on the sum rate. This is because, in an underlay communication network, to prevent primary outage, the interference from the secondary network must be kept under a certain threshold. Hence, when the interference generated by the secondary network reaches this threshold, no more power is allocated for transmission, and beyond this point, increasing $P_{max}$ does not enhance the sum rate. 

However, Fig. \ref{fig1} also shows that there is another way to improve system performance: by increasing the number of T-RIS elements. It can be observed that increasing the number of T-RIS elements results in a higher sum rate for the secondary system for any value of $P_{max}$, and the sum rate of the system converges at a higher rate when the LEO is equipped with a T-RIS having more elements.

\begin{figure}[!t]
\centering
\includegraphics [width=0.48\textwidth]{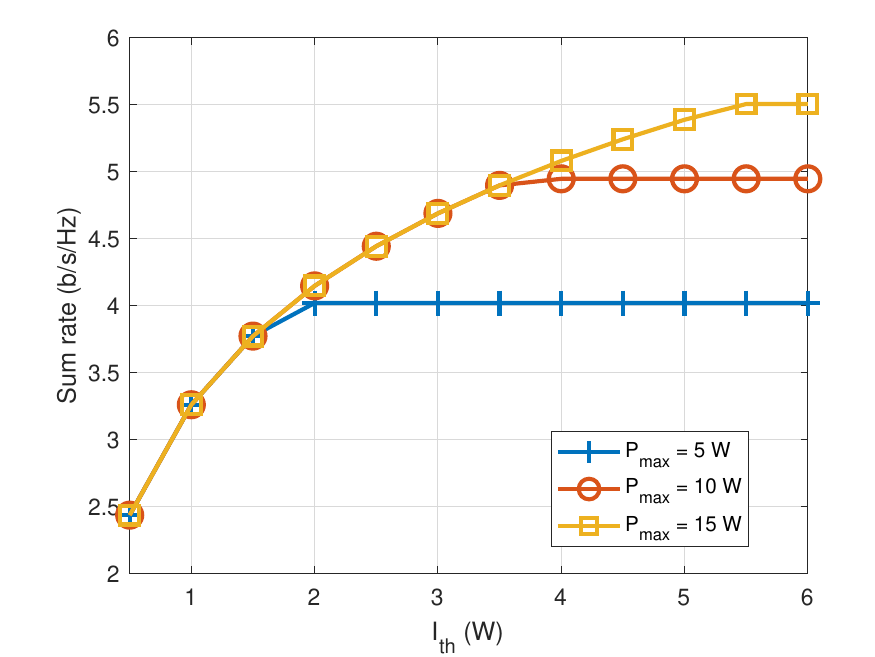}
\caption{Effect of varying $I_{th}$ of the primary network on the sum rate of the secondary network.}
\label{fig2}
\end{figure}

Figure \ref{fig2} shows the effect of increasing the interference threshold of the primary network on the sum rate of the secondary network. When the value of $I_{th}$ is increased, the LEO can allocate more power for transmission without violating the interference threshold, resulting in an increased sum rate for the system. It is also clear from Fig. \ref{fig2} that the sum rate of the system increases with $I_{th}$. However, after a certain point, there is no effect of further increasing $I_{th}$ on the sum rate because, at this point, the LEO is already transmitting with the maximum available power $P_{max}$. Therefore, when the value of $I_{th}$ is increased beyond this point, the LEO cannot transmit with any more power, resulting in a fixed sum rate for the system. Furthermore, it can be seen from the figure that when the value of the available power at the LEO ($P_{max}$) is increased, the sum rate converges at a higher level of $I_{th}$ and at higher values of the sum rate, because in this case, the LEO has more power available for transmission. Therefore, when $I_{th}$ increases, the LEO can increase the transmission power and achieve higher values of sum rates.
\begin{figure}[!t]
\centering
\includegraphics [width=0.48\textwidth]{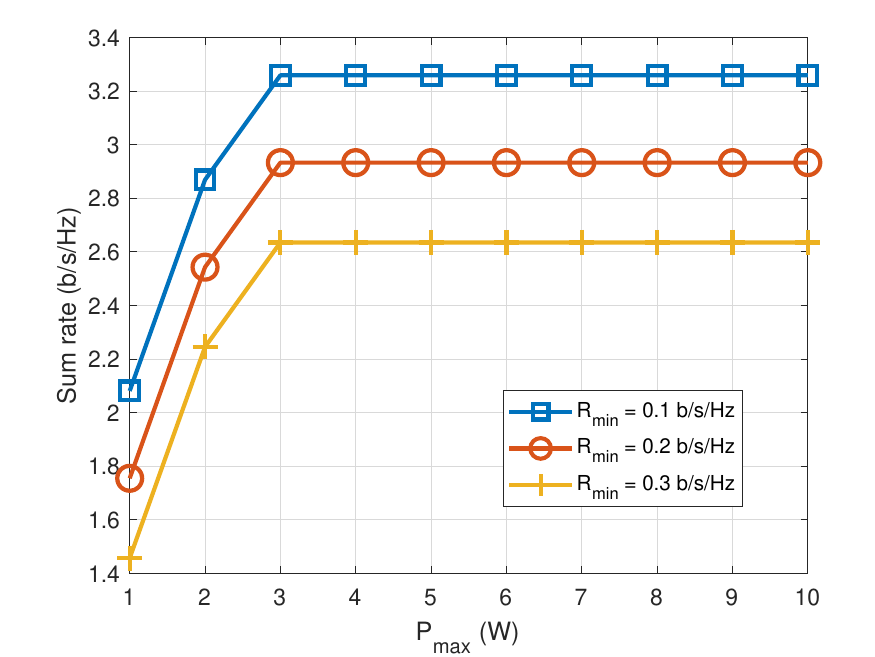}
\caption{The figure illustrates the effect of increasing $P_{max}$ for different values of $R_{min}$.}
\label{fig3}
\end{figure}

Next, Fig. \ref{fig3} shows the effect of different rate requirements ($R_{min}$) of the LIoT and $P_{max}$ on the sum rate of the secondary network. As discussed earlier, higher values of $P_{max}$ result in higher sum rates, and the sum rate converges after a certain point due to the interference threshold. However, it is interesting to see that the system with a smaller value of $R_{min}$ offers higher sum rates compared to the system with larger rate requirements. Even at the points where increasing the value of $P_{max}$ has no effect on the sum rate due to the interference threshold, the system with LIoT having smaller rate requirements offers better performance in terms of the sum rate. This is because, at small values of $R_{min}$, the LIoT with relatively poor channel conditions is allocated transmission power just sufficient to satisfy its minimum rate requirement, and more resources are allocated to the LIoT with better channel conditions to maximize the sum rate of the system. However, as the value of $R_{min}$ increases, the fraction of the available resources allocated for the transmission to the LIoT with poor channel conditions increases, leaving behind a relatively smaller amount of resources for the LIoT with a better channel, resulting in a decrease in the sum rate.

\begin{figure}[!t]
\centering
\includegraphics [width=0.48\textwidth]{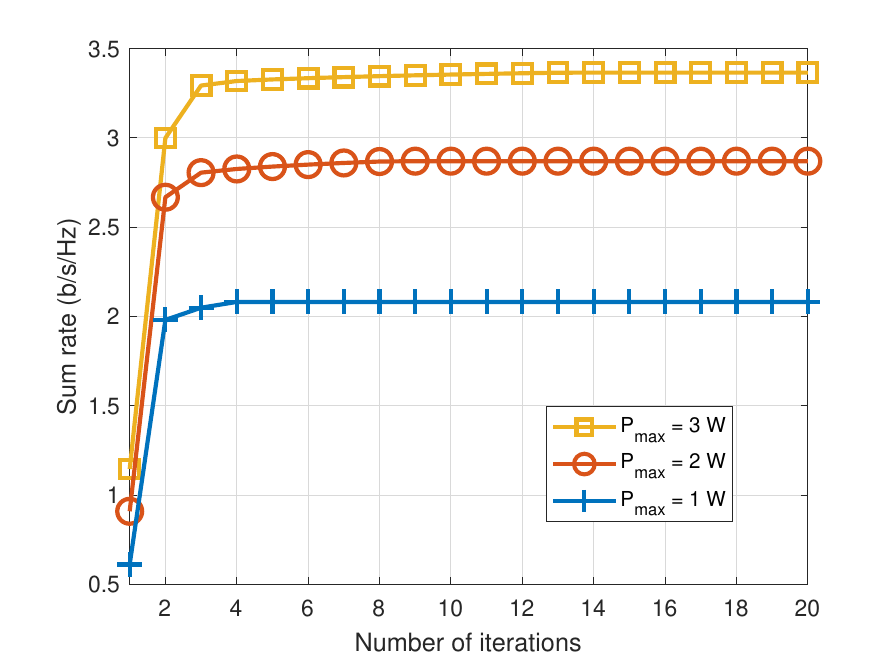}
\caption{The figure shows the convergence behaviour of the proposed framework.}
\label{fig4}
\end{figure}

The convergence behavior of the proposed optimization framework is shown in Fig. \ref{fig4}. It can be seen that the framework provides fast solutions, achieving convergence within 6 iterations, and the values of the available power at the LEO $P_{max}$ have a negligible effect on the convergence rate.
 
\section{Conclusions}
This paper has provided a new optimization framework for CR-enabled NOMA integrated NTNs, where the GEO network operated as a primary network using TDMA and the T-RIS-equipped LEO operated as a secondary IoT network using NOMA transmission. The proposed framework has simultaneously optimized the power allocation of the LEO satellite and the phase shift of T-RIS to maximize the sum rate of the system, subject to the service quality of LIoT and interference temperature from the secondary to the primary network. We first reduced the complexity of the optimization problem using the SCA approach and then implemented the efficient solution in two stages. In the first stage, the power allocation for LEO was calculated using KKT conditions, given the random phase shift of T-RIS. In the second stage, we designed the phase shift of T-RIS using Taylor approximation and semidefinite programming methods, given the optimal power of the LEO satellite. Numerical results demonstrate the effectiveness of the proposed optimization framework.

\bibliographystyle{IEEEtran}
\bibliography{Wali_EE}
\end{document}